# Symbol quantization in interstellar communications:

# methods and observations

*William J. Crilly Jr.*

Green Bank Observatory, West Virginia, USA

*Abstract*— **Interstellar communication transmitters, intended to be discovered and decoded to information bits, are expected to transmit signals that contain message symbols quantized in at least one of the degrees of freedom of the transmitted signal. A hypothesis is proposed that signal quantization, in the form of multiplicative values of one or more signal measurements, may be observable during the reception of hypothetical discoverable interstellar communication signals. In previous work, using single and multiple synchronized radio telescopes, candidate hypothetical interstellar communication signals comprising $\Delta t$ $\Delta f$ opposite circular polarized pulse pairs have been reported and analyzed. (ref. arXiv:2105.03727, arXiv:2106.10168, arXiv:2202.12791). In the latter report, an apparent quantization of $\Delta f$ at multiples of 58.575 Hz was observed. In the current work, a machine process has been implemented to further examine anomalous $\Delta f$ and $\Delta t$ quantization, with results reported in this paper. As in some past work, a 26 foot diameter radio telescope with fixed azimuth and elevation pointing is used to enable a Right Ascension filter to measure signals associated with a celestial direction of interest, relative to other directions, over a 6.3 hour range of Right Ascension. The $5.25 \pm 0.15$ hr Right Ascension, $-7.6° \pm 1°$ Declination celestial direction presents repetition and quantization anomalies, during an experiment lasting 157 days, with the first 143 days overlapping the previous experiment.**

## I. INTRODUCTION

Communication signals may be classified by the ease with which transmitted signals may be discovered, received and decoded to information bits. At opposite extremes of classification, transmitted signals may be designed to be almost indistinguishable from random noise, or, may continuously present a single orthonormal function, or single measurement value, in each degree of freedom [1][2]. Examples of extremes include spread spectrum modulated signals, and unmodulated carriers. The former provides high information rate at the expense of discovery, while the latter provides discovery at the expense of information rate. Interstellar communication systems are hypothesized to present receiver measurements between these extremes.

An additional expectation of interstellar signals is that signal measurements should themselves provide a logical path to at least some signal message decoding. A single isolated anomaly should not require an increase in receiver sensitivity, to further examine the signal. For example, a signal's carrier may be used to measure the center frequency and arrival direction of a signal. However, its transmitted power is wasted if the highest sensitivity receiver on a planet discerns no other signal anomaly. The discovery of an anomalous signal measurement should preferably present at least one additional measurement that leads the same receiver to glean a process to decode the signal to information bits, or to an analog representation of information. A signal comprising opposite circular polarized pulse pairs is an example of such discoverable signals.

In the current work, hypothetical quantization of the difference in measured frequency of opposite circular polarized pulses is examined, while assuming that an Additive White Gaussian Noise (AWGN) model, augmented with Radio Frequency Interference (RFI) amelioration methods, is expected to explain measurements.

The remainder of this paper contains a hypothesis, method of measurement, observations during a 157 day experiment, discussion and conclusions.

## II. HYPOTHESIS

*Hypothesis*: Hypothetical energy-efficient discoverable interstellar communication signals [3][4], comprising $\Delta t$ $\Delta f$ opposite circular polarized pulse pairs, are expected to present quantization and repetition of one or more of the degrees of freedom of the signal. The hypothesis may be conditionally falsified [5] by finding an absence of such radio telescope receiver measurements, assuming an RFI-augmented AWGN signal explanatory model. Measurements are to be diurnally performed across a range of Right Ascension (*RA*) values, including a prior celestial direction of interest at $5.25 \pm 0.15$ hr *RA*, $-7.6° \pm 1°$ Declination (*DEC*).

## III. Method of Measurement

Traceability, repeatability and testability are primary objectives of the measurement protocol in this work. Consequently, the method of measurement is entirely machine implemented, from radio telescope signal capture to presentation of results image production. A few anecdotal insights were used in the establishment of hyperparameters in the quantization filters in the present work, and are described in the paragraph below. Further anecdotal insight is often most useful after a methodical procedure has been implemented. Then, a collection of anecdotal insights may guide and prioritize further work.

The methods described in previous reports [6][7][8] are retained in the current work, with an additional process to build and test two $\Delta f$ multiplicative quantization filters, processing previously captured files, and new radio telescope files, comprising 153 days of 100% duty-cycle dual circular polarized radio telescope data. The $\Delta f$ quantization filters added in the current work quantize $|\Delta f|$ to 58.575 Hz multiples with ± 10.5 Hz span, and half these values, using $|\Delta f|$ / quantized to 29.288 Hz multiples spanning ± 5.5 Hz. In other words, in the 58.575 Hz quantizing filter, the quantized filtered $|\Delta f|$ range of 80 to 400 Hz, is reduced to output pulse pairs' data that have $|\Delta f|$ values that measure within 10.5 Hz of multiples of 58.575 Hz. The 58.575 Hz and ± 10.5 Hz range were set from the 3.7 Hz FFT bin width, equipment metrology estimates, and the observation of high SNR$_{METRIC}$ anomalous $\Delta t$ = -3.75 s polarized pulse pairs. The 29.288 Hz ± 5.5 Hz quantization filter values are chosen to be close to half of the 58.575 Hz and ± 10.5 Hz values.

During exploratory investigation of quantization hyperparameters, an extended range $\Delta f$ filter is implemented, with results reported in **Figure 11**. The use of various quantization filters is a topic of ongoing and future work.

## IV. Observations

**Figures 1 through 11** present image files produced by machine processes of radio telescope data files, in a multi-step signal processing system. The machine processes, measured parameters, and machine hyperparameters are described in previous reports [6][7][8], in Method of Measurement above, and in this section.

The experiment conducted in recent work spanned 157 days, with 153 days of telescope data included in the presentation of results. Days not included were the result of power outages and equipment issues. Due to the 143 day overlap of measurement days in the quantized filter results presented in the current report, and unquantized filter output reports in [8], there will be similarities seen between the sets of plots in this work and in [8].

**Figure 1** plots $\Delta t$ = -6.25 s, 29.288 Hz quantized polarized pulse pairs, against $RA$ bins from 0 to 6.3 hours in 0.3 hour intervals. The 29.288 Hz filter was chosen after the twice valued 58.575 Hz quantized filter was gleaned from post-processed measurement results reported in [8] Figure 6.

**Figure 2** plots the Modified Julian Date (MJD) of the polarized pulse pair events observed in **Figure 1**, i.e. of a $\Delta t$ = -6.25 s $\Delta f$ quantized measurement set.

**Figure 3** plots the RF Frequency measurement of the polarized pulse pair events of the $\Delta t$ = -6.25 s $\Delta f$ quantized measurements. Widely spaced RF frequencies are expected from a transmitter intending the receiver to conduct Angle of Arrival (AOA) measurements, described in Transmitter Design, and Further Work in [6]. In addition, communication information rate and RFI amelioration are improved. Ideas about the possible presence of relatively narrow bandwidth energy bursts, spread across a wider bandwidth, are detailed in Discussion in [8].

**Figure 4** plots the $\Delta f$ of the polarized pulse pair events, of the $\Delta t$ = -6.25 s $\Delta f$ quantized measurement set.

**Figure 5** plots the log likelihood of the $\Delta t$ = -3.75 s events observed during the 157 day duration experiment, after quantizing the $|\Delta f|$ to be within 10.5 Hz of a multiple of 58.575 Hz and within a range of 80 to 400 Hz.

**Figure 6** plots the MJD of the $\Delta t$ = -3.75 s quantized $\Delta f$ events plotted in **Figure 5**.

**Figure 7** plots RF Frequency measurement of the $\Delta t$ = -3.75 s quantized $\Delta f$ events plotted in **Figure 5**. RF Frequency is spread, for possible reasons described above in **Figure 3**.

**Figure 8** plots the $\Delta f$ measurement of the $\Delta t$ = -3.75 s quantized $\Delta f$ events plotted in **Figure 5**.

**Figure 9** plots log likelihood of polarized pulse pairs in the 5.1 – 5.4 hr $RA$ range, over $|\Delta t| \leq 10$ s using the described 58.575 Hz quantizing filter. Multiple anomalous polarized pulse pairs are presented, given that one polarized pulse pair is expected below a log likelihood of -2.0, per measurement.

**Figure 10** plots log likelihood of polarized pulse pairs in the 5.1 – 5.4 hr $RA$ range, over $|\Delta t| \leq 10$ s using the 29.288 quantizing filter. As observed in **Figure 9**, multiple anomalous polarized pulse pairs are presented below a log likelihood of -2.0. Certain $|\Delta t|$ values matching in **Figure 9** and **Figure 10** are described in text below the two figures.

**Figure 11** shows results of a high SNR$_{METRIC}$ polarized pulse pair exploratory search above the 400 Hz previous upper limit of $|\Delta f|$. An unusual set of five pulse pairs were observed at $\Delta t$ = +7.25 s, in the 5.1 to 5.4 hr $RA$ direction. The five pulse pairs were observed on MJD 59588, 59517, 59575, 59515, and 59592. The $\Delta f$ quantized value range threshold, set to ±3.5 Hz, is lower than the value used in other observations. One possible explanation for the reduced range entails the possibility that all of the five pulse pairs computed no difference in post-multiplied FFT bin index, relative to quantizer multiplier predicted bin indexes, and receiver equipment metrology did not increase this value above one FFT bin width, i.e. 3.725 Hz. Intentionally discoverable transmitters sometimes provide a means for the receiving entity to verify that receiver processing is working as intended. The receiver test then occurs as a natural result of transmitting a combination of repetitive and quantized transmitter settings, within discoverable symbols. There is a possibility that a transmitter is designed with this intended purpose.

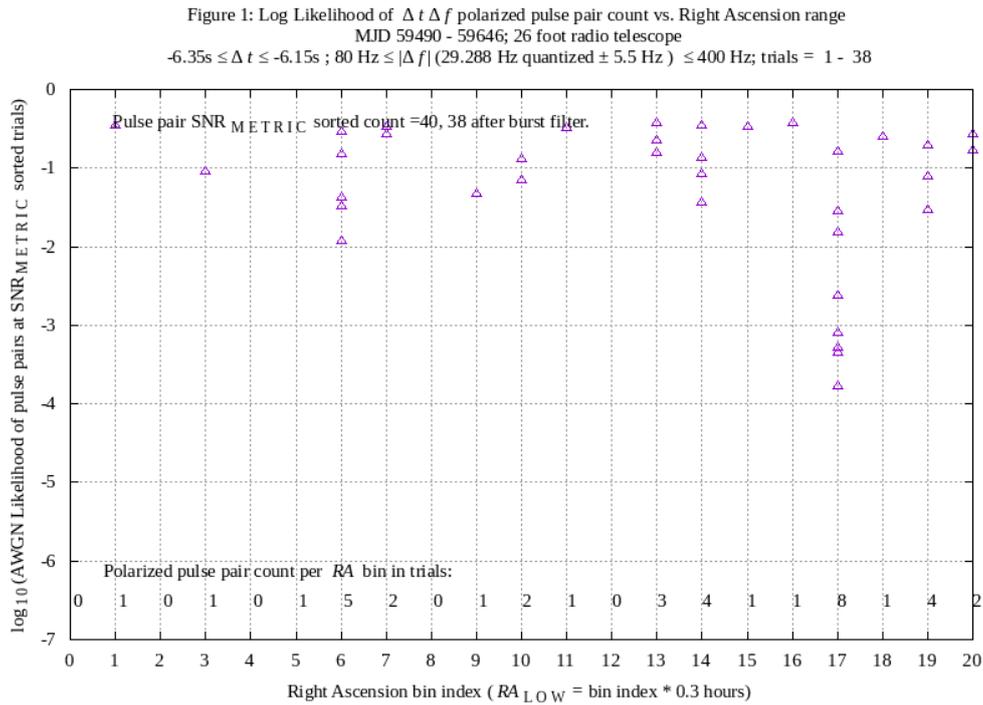

**Figure 1:** 29.288 Hz $\Delta f$ quantized polarized pulse pairs having $\Delta t = -6.25$ s, observed in the 5.1 to 5.4 hour $RA$ range, during 157 days, present likelihood approximately a thousand times less than expected from an RFI-augmented AWGN noise model. The binomial distribution of eight events seen in 38 tries, given an event probability of 1/21, equals 3 x $10^{-4}$. The highest five $SNR_{METRIC}$ $RA$ bin 17 $\Delta f$ quantized polarized pulse pairs are ranked 4 ,6, 13, 14 and 16.

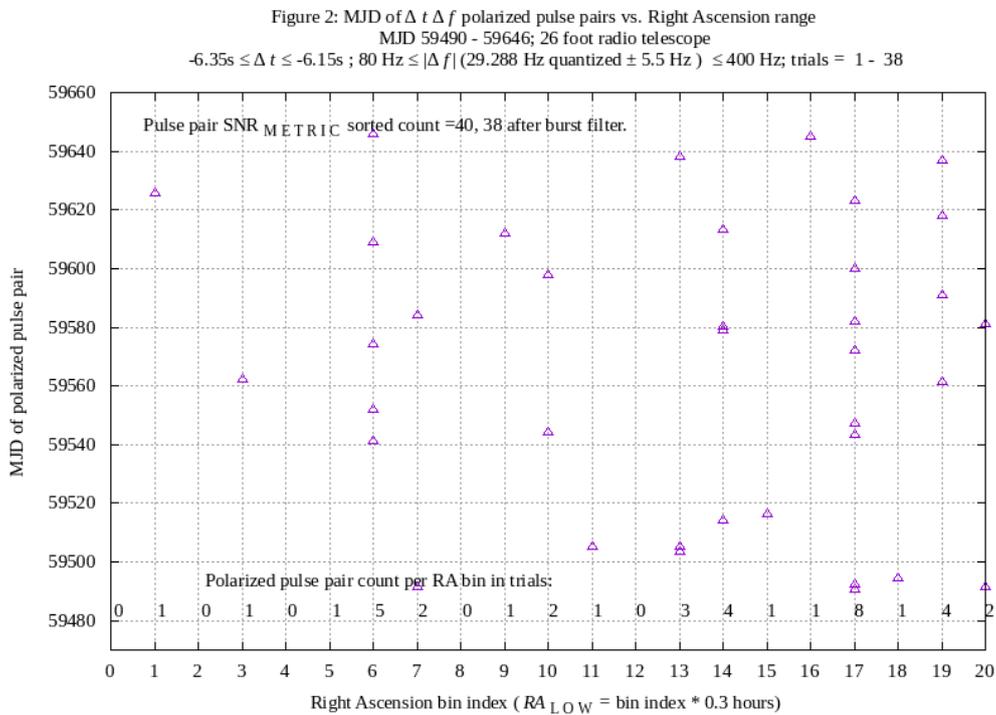

**Figure 2:** The MJDs of the eight quantized polarized pairs observed in **Figure 1** appear distributed, potentially refuting a hypothesis of some diurnal RFI models explaining the events. Events do not appear to be repeatedly replicated on the same MJD, expected from intra-day persistent RFI. Robust RFI filters, quantization filters and $RA$ binning reduce the number of observed events to less than or equal to one, in thirteen of the twenty-one $RA$ bins, implying that RFI may have been largely removed using machine processes.

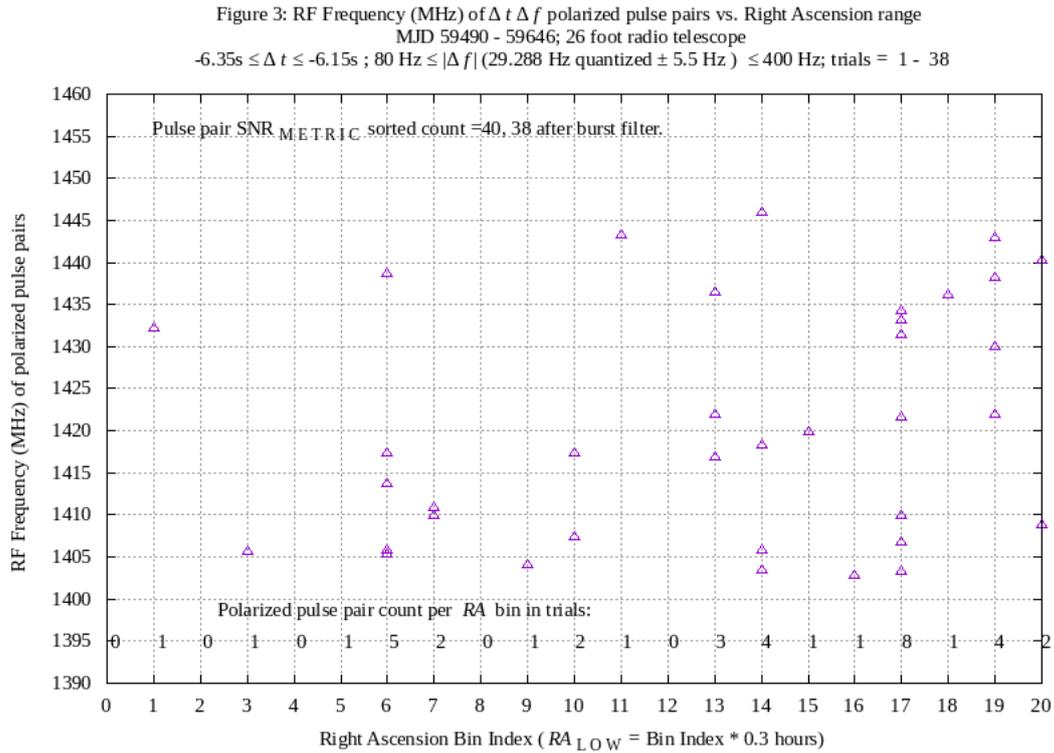

**Figure 3:** A high concentration of RF frequency of filtered quantized polarized pulse pairs does not appear evident. There might be significance to a higher number of pulse pairs observed at lower RF frequency.

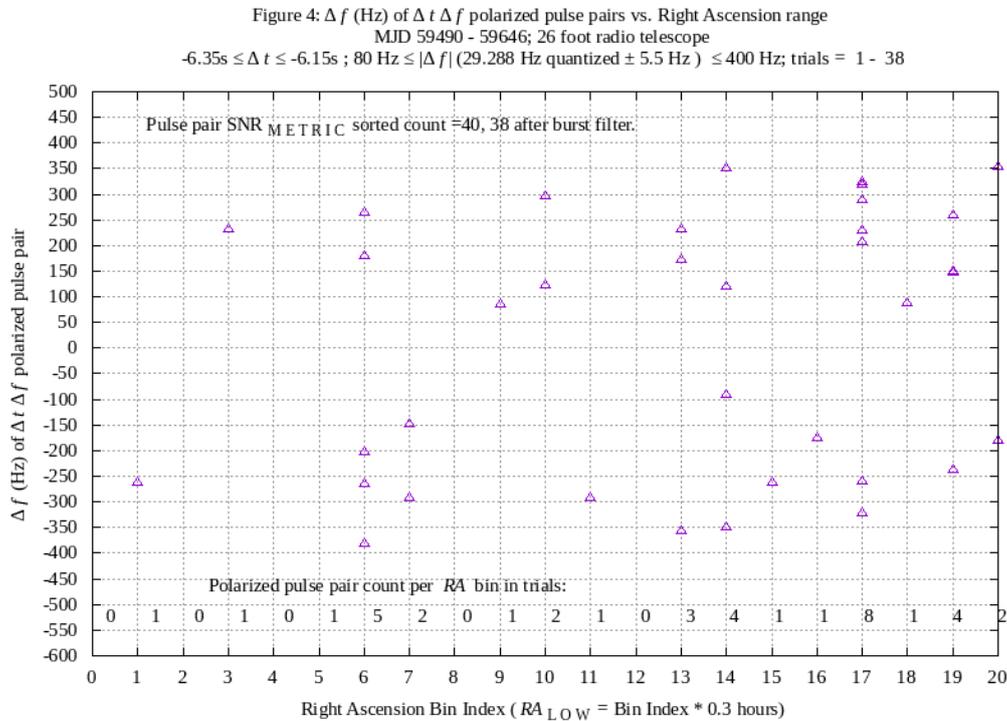

**Figure 4:** $|\Delta f|$ of apparently significant bin 17 pulse pairs may be concentrated to a region less than the 80 to 400 Hz range, particularly at the low range. The low range of $|\Delta f|$ rejection is useful in the receiver to ameliorate narrowband RFI that may mimic pulse pairs, due to Doppler spread of an RFI source-to-radio telescope propagation path.

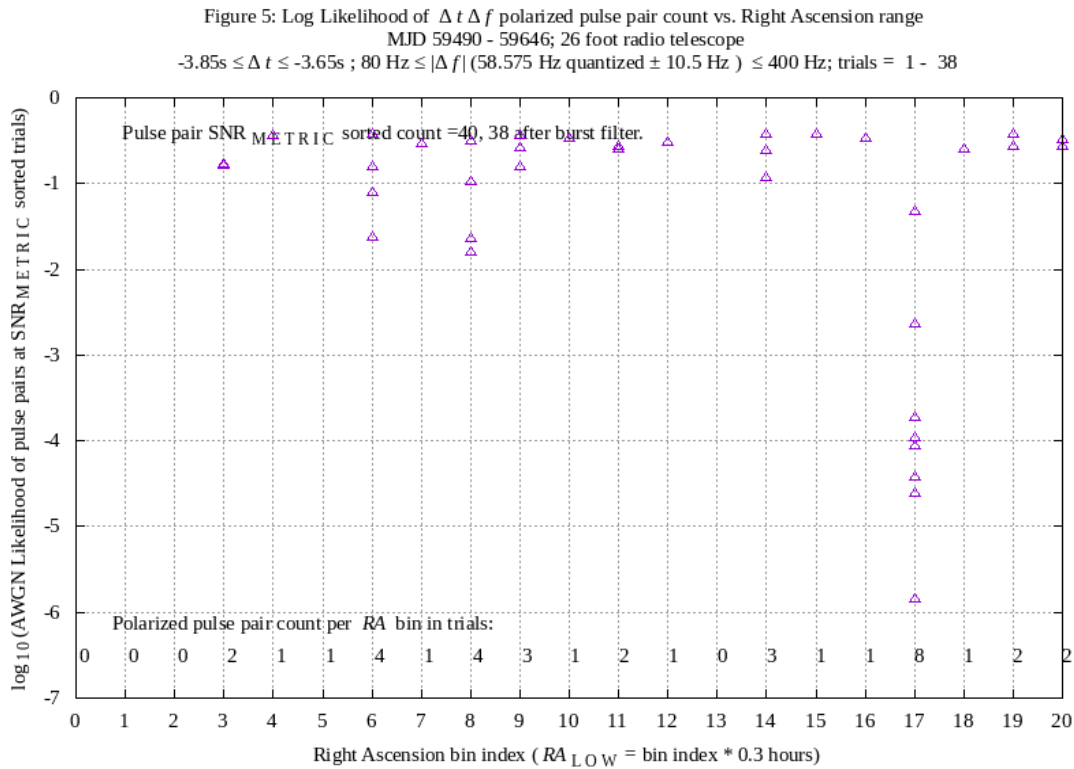

**Figure 5:** 58.575 Hz $|\Delta f|$ quantized polarized pulse pairs having $\Delta t = -3.75$ s, observed in the 5.1 to 5.4 hour *RA* range, during a 157 day experiment, present likelihood approximately a hundred thousand times less than expected from an RFI-augmented AWGN noise model.

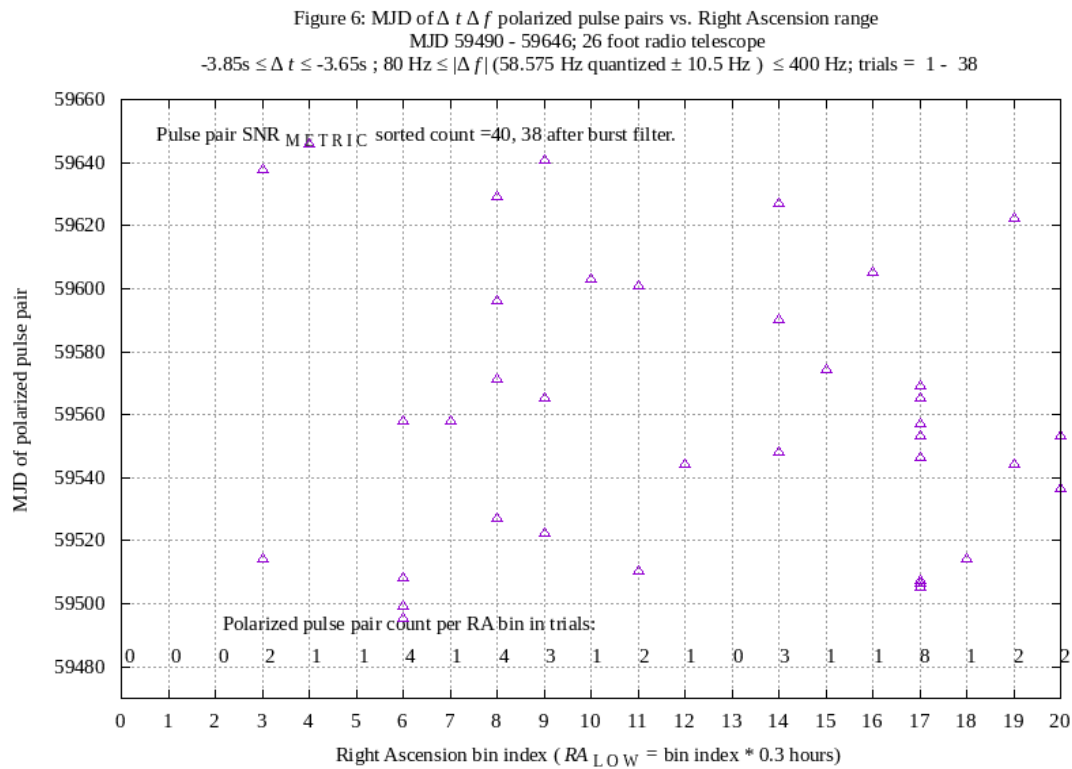

**Figure 6:** $\Delta f$ quantized polarized pulse pairs observed in the 5.1 to 5.4 hour *RA* range appear to present an MJD concentration greater than concentrations presented in other *RA* ranges. An intentionally discoverable transmitter might be designed to concentrate highest power transmissions during a group of days.

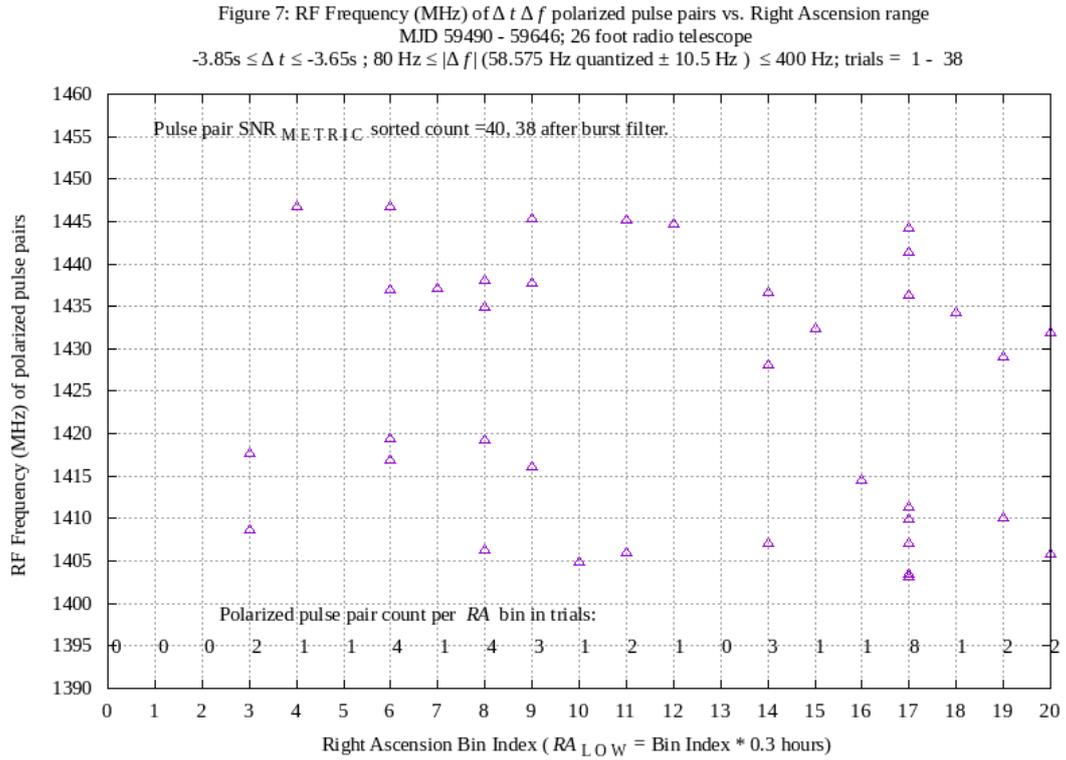

**Figure 7:** RF frequency of quantized polarized pulse pairs in the direction of interest appear distributed, with an exception near 1403 MHz.

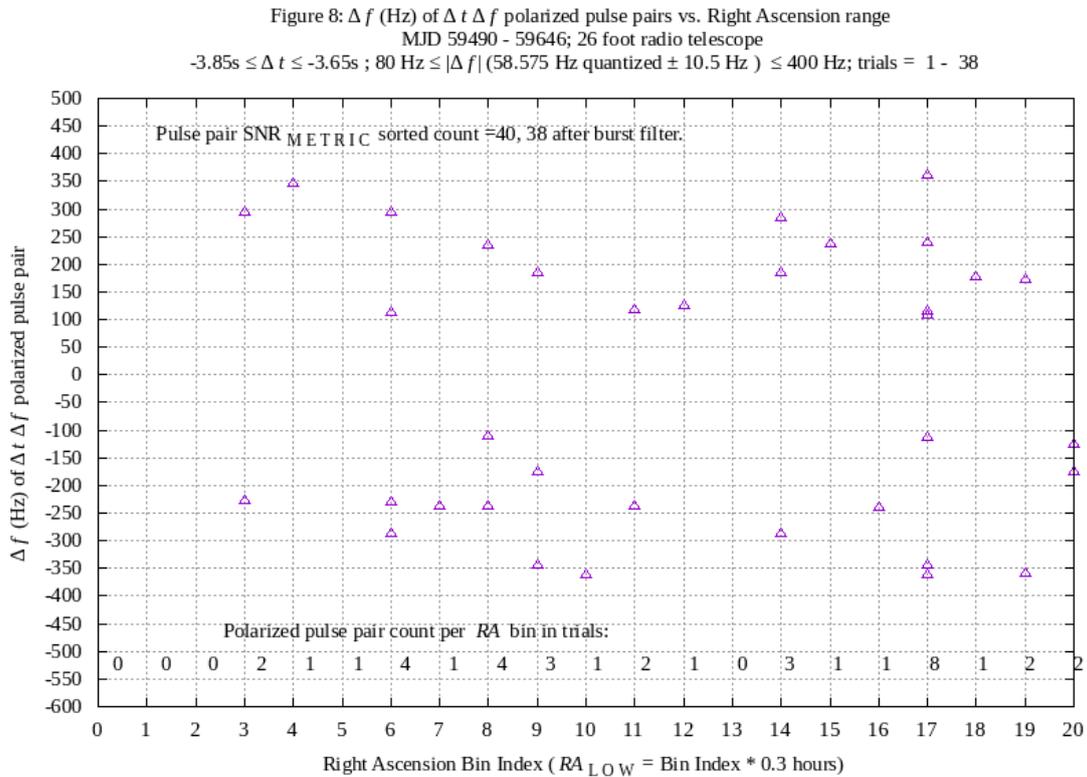

**Figure 8:** $\Delta f$ measurements of polarized pulse pairs is plotted for the $\Delta t$ = -3.75 s apparently repetitive and quantized measurements, described in prior work, e.g. Figure 6 of [8]. Comparison of the bin 17 indications implies that the quantization filter significantly reduces the filter output response in other *RA* bins, relative to bin 17, 5.1 − 5.4 hr RA.

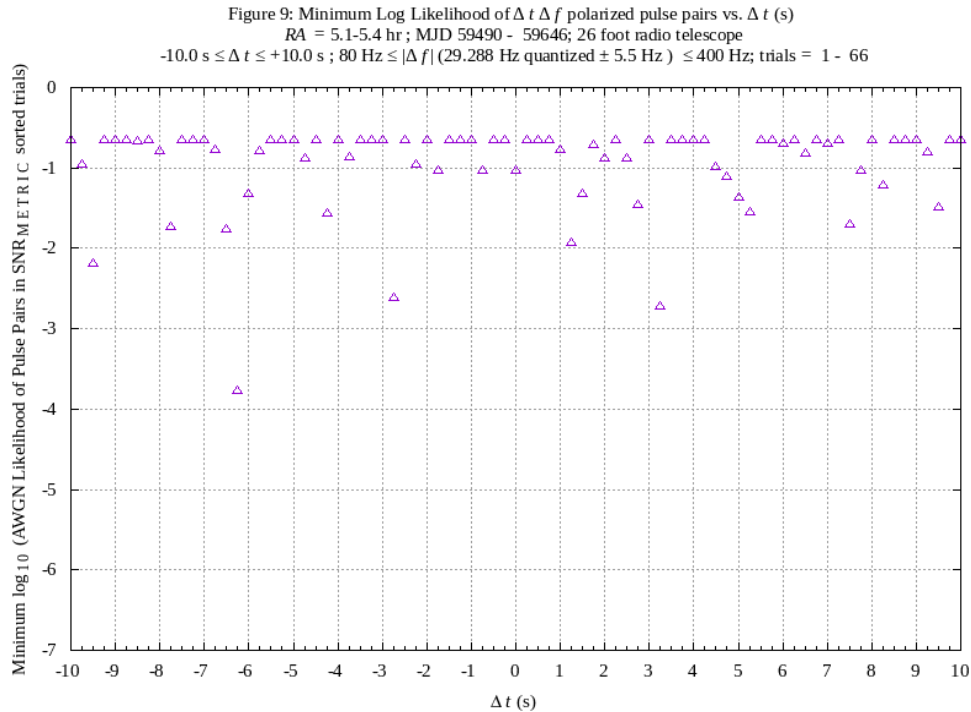

**Figure 9:** The process used to measure log likelihood of quantized repetitive $\Delta t$ symbol count in *RA* ranges, yielding **Figure 1** and **Figure 5**, was modified to measure minimum log likelihood in the 5.1 to 5.4 hr *RA* range, at 81 $\Delta t$ values, -10.0 $\leq \Delta t \leq$ 10.0 s. Several settings of $\Delta t$ indicate anomalous log likelihood, i.e. at $\Delta t$ values of -9.5 s, -6.25 s, -2.75 s, +1.25 s and +3.25 s. Log likelihood, due to an RFI-augmented AWGN model, is estimated to occur with one log likelihood event having a value less than -2.0, per 157 day duration experiment. Four such events are observed.

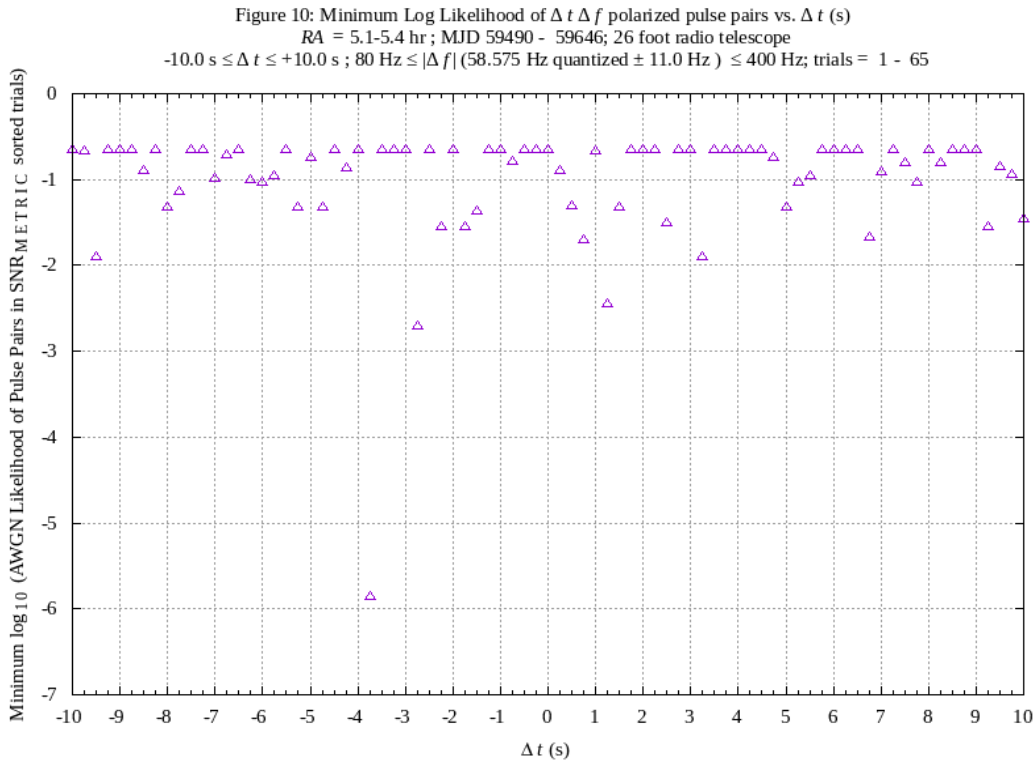

**Figure 10:** In addition to 58.575 Hz quantized anomalies observed at $\Delta t$ = -3.75 s, four $\Delta t$ values -9.5s, -2.75 s, +1.25 s and +3.25 s, observed in **Figure 9,** appear significant at 58.575 Hz quantization. Each of the four $\Delta t$ events is present, using an RFI-augmented AWGN model probability, at $\leq$ -1.9 log likelihood. The binomial distribution of five events seen in 81 trials, at event pr. = $10^{-1.9}$, equals 0.003. The presence of four of the same $\Delta t$ values observed in **Figure 9** data, supports a thought that $\Delta t$ and $\Delta f$ in polarized pulse pairs are each quantized and repetitive.

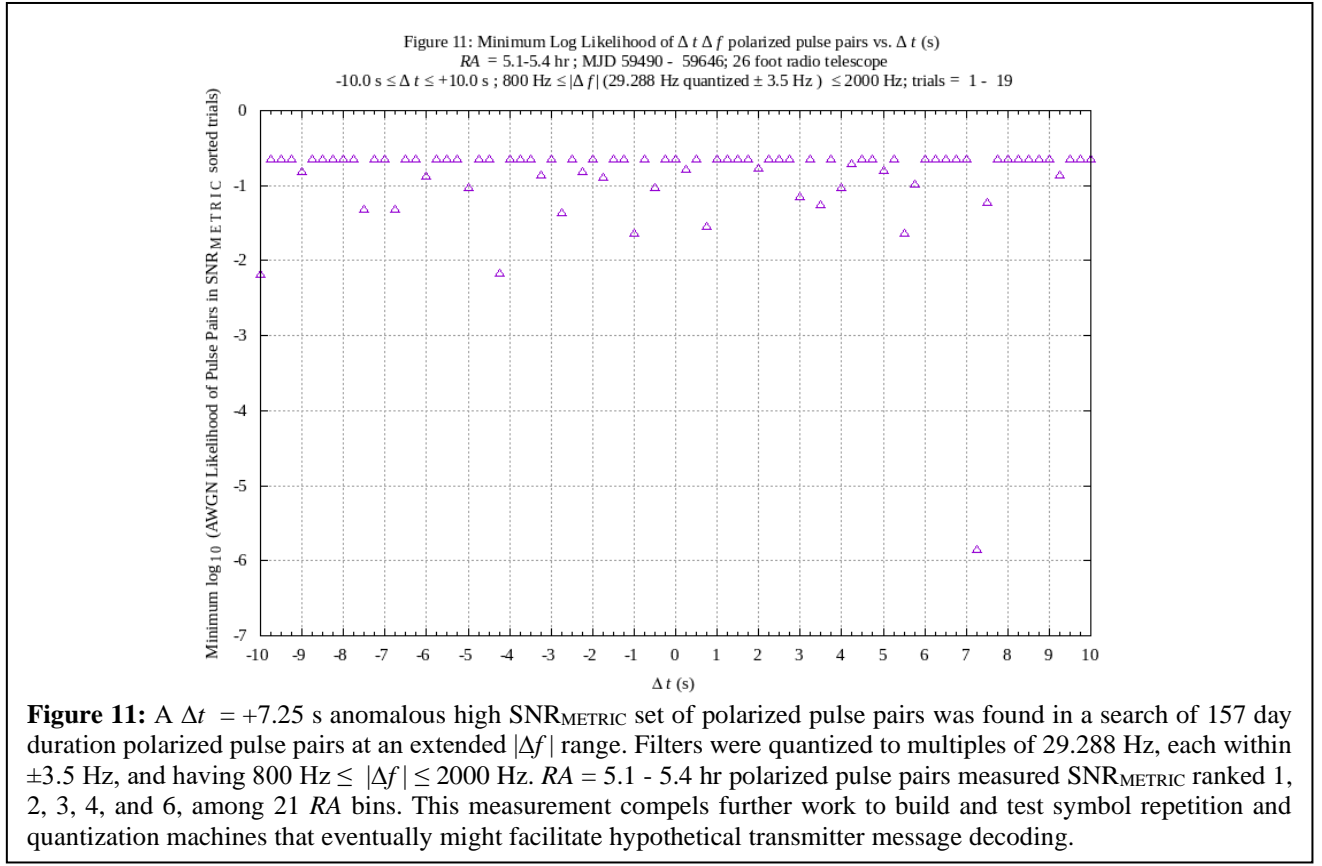

**Figure 11:** A $\Delta t = +7.25$ s anomalous high SNR$_{\text{METRIC}}$ set of polarized pulse pairs was found in a search of 157 day duration polarized pulse pairs at an extended $|\Delta f|$ range. Filters were quantized to multiples of 29.288 Hz, each within $\pm 3.5$ Hz, and having 800 Hz $\leq |\Delta f| \leq 2000$ Hz. $RA = 5.1 - 5.4$ hr polarized pulse pairs measured SNR$_{\text{METRIC}}$ ranked 1, 2, 3, 4, and 6, among 21 $RA$ bins. This measurement compels further work to build and test symbol repetition and quantization machines that eventually might facilitate hypothetical transmitter message decoding.

## V. Discussion

Many explanations may help one understand the cause of observed anomalies. Ideas and examples of explanations are described below.

**Potential population selection bias**

The observed 58.575 Hz, and half of 58.575 Hz apparent quantization of $\Delta f$ in polarized pulse pairs, may perhaps be explained using an auxiliary hypothesis that predicts the choice of these value to be statistical noise induced by population selection bias. The 58.575 Hz value was chosen due to investigation of $\Delta t = -3.75$ s repetitive pulse pairs reported in [8] Figure 6. The 29.288 Hz half value was chosen during examination of the RFI-augmented AWGN model's log likelihood of the apparently repeating $\Delta t = -6.25$ s polarized pulse pairs. When a quantization base value is chosen from the results of an experiment, one requires models and statistical analysis to test significance of the chosen value, e.g. using the $pr. = 0.05$ composite value calculated from observations described in [8] Observations, related to [8] Figure 6. On the other hand, observation of twice harmonic quantization base values, i.e. at 29.288 Hz and 58.575 Hz, on almost alternate sets of days, implies a more complex auxiliary hypothesis, describing selection bias, that is yet to be developed.

**Equipment issues**

Anomalies may be caused by equipment issues. Some possibilities are described.

1. The use of numerous RFI rejection filters [6][7][8], each having chosen hyperparameters, e.g. filters, rejection thresholds, risks a leakage of RFI into measurement results. RFI algorithm output files need to be examined.

2. Telescope data capture processes may lead to RFI leakage into measurement results. Simulations and further long term artificial noise source tests are required to understand this possibility.

3. Natural objects, together with equipment algorithms, may produce results that mimic communication signals.

4. The physical temperature of the low noise amplifiers at the telescope feed are expected to induce receiver sensitivity changes over time, affecting measurements.

5. Software errors may be present.

**Absence of corroboration increases uncertainty**

At present, independent observation and analysis of polarized pulse pairs in the hypothetical direction of interest are not known to have been performed. Sporadic presence and pointing uncertainties of the reported highest SNR pulse pairs make follow-up with larger single pixel radio telescopes potentially time consuming. Prediction of pulse pair events and direction-finding seems important to ameliorate the follow-up issue.

**Limited, or changing, matched filter range**

Apparent repetition and quantization of signal measurements compels a search for potentially associated anomalies to help explain measurement results. An exploratory search for pulse pairs having $|\Delta f|$ greater than 400 Hz has been conducted, with some preliminary indication of anomalous results above 400 Hz, plotted in

Figure 11. There is a possibility that a transmitter is switching its parameters over time, to allow greater likelihood of discovery, a lower risk of mimicking RFI, and/or given thoughts that a receiver might be set to a relatively confined and fixed set of filters for a long duration, for reasons unknown to the transmitter. Switching transmitter parameters from one set to another may be an inherent aspect of discoverable interstellar communications transmitters. Electromagnetic propagation systems that are designed to be difficult to detect are expected to not repeat quantization base values during relatively short time periods. RADAR systems, and secure communications systems often develop a variety of methods to prevent discovery. These concealment methods do not appear to be present in observations in this and previous work. Polarized pulse pairs are observed using wide beamwidth, wide bandwidth radio telescopes without difficult searches.

**Correlations may be difficult to explain**

The apparent correlation of high $SNR_{METRIC}$ levels, having multiples of a base $\Delta f$ value, at one $\Delta t$ value, and also at multiples of half the base $\Delta f$ value, at a different $\Delta t$ value, and having $RA$ correlation during long periods, seems difficult to explain, without introducing various intentional transmitter models and unusual RFI models. On the other hand, there may be an ultimately simple explanation currently unknown.

## VI. CONCLUSIONS

The RFI-augmented AWGN model continues to not explain experimental results. Observations indicate that repetition and quantization of polarized pulse pair measurements seem to exist, associated with the hypothetical celestial direction of interest. However, a definitive conclusion is not presently possible, due to many other possible explanations. The mechanisms that produce the anomalies observed in this work are therefore concluded to be not understood, compelling further work.

## VII. FURTHER WORK

Items of further work have been proposed in past reports [6][7][8]. Prioritization should focus on better understanding the anomalies that are being currently observed. Given this idea, a search for additional repetition, quantization, and hyperquantization [9] seems important, together with the development of polarized pulse pair prediction, squelch and direction-finding equipment algorithms.

## VIII. ACKNOWLEDGEMENTS